\documentclass[aps,pre,reprint,groupedaddress,showpacs]{revtex4-1}

\usepackage{amsmath,graphicx}

\begin{document}

\title{Electron mean free path from angle-dependent photoelectron spectroscopy of aerosol particles}

\author{Maximilian~Goldmann}
\author{Javier~Miguel-S\'anchez}
\author{Adam~H.~C.~West}
\author{Bruce~L.~Yoder}
\author{Ruth~Signorell}
\email{rsignorell@ethz.ch}
\affiliation{ETH~Z\"urich, Laboratory of Physical Chemistry, Vladimir-Prelog-Weg 2, 8093~Z\"urich, Switzerland}

\date{\today}

\begin{abstract}
We propose angle-resolved photoelectron spectroscopy of aerosol particles as an alternative way to determine the electron mean free path of low energy electrons in solid and liquid materials. The mean free path is obtained from fits of simulated photoemission images to experimental ones over a broad range of different aerosol particle sizes. The principal advantage of the aerosol approach is twofold. Firstly, aerosol photoemission studies can be performed for many different materials, including liquids. Secondly, the size-dependent anisotropy of the photoelectrons can be exploited in addition to size-dependent changes in their kinetic energy. These finite size effects depend in different ways on the mean free path and thus provide more information on the mean free path than corresponding liquid jet, thin film, or bulk data. The present contribution is a proof of principle employing a simple model for the photoemission of electrons and preliminary experimental data for potassium chloride aerosol particles.
\end{abstract}


\maketitle

\section{Introduction\label{intro}}

The inelastic and elastic mean free path (MFP) of an electron is the average distance an electron of a given kinetic energy travels in a material between two successive inelastic or elastic collisions, respectively~\cite{Chen1992,Jablonski1999}. Electron mean free paths (EMFPs) are important quantities for a number of physico-chemical phenomena. These range from aerosol physics and chemistry, which influence the radiative energy balance of the Earth and other planets, to photoemission processes in interstellar dust clouds, radiotherapy, and nuclear plant operation (see~\cite{Burtscher1982,Ziemann1998,Wilson2006,Signorell2010,Young2012,Litman2013,Zhang2013,Signorell2014,West2015,Winter2006,Yamamoto2014,Gail1980,Watson1972,Gervais2006,Gu2012,Alizadeh2012} and references therein). A complete understanding of the whole photoemission process from the absorption of light to the emission of electrons from the material is crucial in this context~\cite{Hertz1887,Margaritondo1988,Chiang2001}. For many compounds, the values of the inelastic and elastic MFP are not known or inaccurate, largely due to various experimental difficulties. Especially challenging are experiments at electron kinetic energies (KEs) below 100~eV and for liquids, which cannot be brought into vacuum as bulk~\cite{Tanuma1988,Tanuma1991,Akkerman1996,Powell1974,Michaud1987}. While several methods have been introduced for the energy range from 15~eV to 100~eV, the uncertainties of the mean free paths remain high in the range below 15~eV~\cite{Michaud1987,Bourke2010}. Recently, progress has also been made for liquids by employing liquid microjets~\cite{Thurmer2013,Suzuki2014}.

Here, we introduce an alternative way to determine the MFPs of slow electrons with KEs below 20~eV which relies on angle-resolved photoelectron spectroscopy of aerosol particles. Aerosol particles are ionized by single photons from a vacuum ultraviolet (VUV) light source. The three-dimensional (3D) distribution of the generated photoelectrons is then projected with a velocity map imaging (VMI) setup ~\cite{Eppink1997} on a two-dimensional (2D) position sensitive electron detector~\cite{Yoder2013,Signorell2014} (section~\ref {exp}). To characterize the properties of photoelectrons, we extract four parameters from the photoelectron images in addition to the electron kinetic energy: the total electron yield and three anisotropy parameters, which describe the anisotropy of the photoelectrons with respect to the propagation and the polarization direction of the ionizing radiation. These four parameters depend in a characteristic way on the EMFPs and the aerosol particle size. The EMFPs are extracted from fits of the simulated anisotropy parameters to the experimental ones varying the EMFPs. The simulated photoemission images are obtained from a simple model (section~\ref {model}). Similar models have been widely used in the past, also for particulate matter (see~\cite{Bohren1983,Yurkin2007,Watson1973,Abbati1974,Muller1988,Shimizu1992,Berg2012} and references therein). Briefly, classical electrodynamics is employed to treat light-particle interaction and a simple three-step model is used to describe electron formation, transport in, and emission from the aerosol particles.

The use of the aerosol approach has two main advantages compared with other methods. Firstly, it allows one to determine EMFPs also for liquids, which cannot be brought into vacuum as thin films or bulk due to their high vapor pressure. This is not a fundamental problem for aerosol droplets. Secondly, aerosol data recorded over a range of different aerosol particle sizes in principle contain more information on the EMFPs than corresponding thin film, liquid jet, or bulk data, because the size-dependence of the anisotropy can be exploited in addition to the kinetic energy. The key point is as follows: The variation of the aerosol particle size changes the total path length of the ionizing radiation and the total scattering path lengths of the electrons inside the particles. In addition, the variation of size corresponds to a variation in curvature. This changes the geometry of electron escape from the particle. More importantly, the curvature characteristically modifies the pattern of the light intensity inside the particle and thus the ionization probability. In other words, the curvature strongly influences the anisotropy of the photoelectrons. Since all these finite size effects depend in different ways on the EMFP they provide characteristic and complementary information on the EMFP.

The purpose of this contribution is a proof of principle of the idea. We use potassium chloride (KCl) as a model substance because many of its properties are well known and aerosol formation is straightforward~\cite{Roessler1968,Wertheim1995,Ejiri1994,Hansson1998}. Furthermore, calculations for EMFPs at low KEs already exist for KCl~\cite{Akkerman1994}, which allow for a comparison with results from the present work. In section~\ref{resu}, we first present the main prediction from modeling, which we then compare with preliminary experimental data for KCl aerosol particles.

\section{Model and parameter definition\label{model}}

The model consists of three parts: The \emph{photonic model} (subsection~\ref{model-1}) describes the interaction of the aerosol particle with the electromagnetic radiation. The \emph{electron emission model} (subsection~\ref{model-2}) includes the formation of the electron at a given location inside the aerosol particle, the transport of the electron through the particle, and its emission at the particle surface. The \emph{electron projection model} (subsection~\ref{model-3}) defines the projection of the 3D photoelectron distribution onto a 2D detector plane. This subsection also introduces the parameters, $\alpha$, $\beta_{\text{backward}}$, $\beta_{\text{forward}}$, and $\gamma$, that describe the characteristics of the photoemission for comparison with experimental data. We use a 3D Cartesian coordinate system with the origin at the center of the aerosol particle. The photons travel in the positive $x$~direction and are linearly polarized along the $y$ axis. The aerosol particles travel in the negative $y$~direction and the photoelectrons are projected in the positive $z$~direction onto the 2D detector plane which is parallel to the $x$-$y$ plane.

\subsection{Photonic model\label{model-1}}

Classical continuum electrodynamics is employed for modeling light-particle interaction. For the KCl particles, we use the frequency-dependent refractive index data in the VUV from Ref.~\cite{Roessler1968} and assume spherical particles surrounded by vacuum. Note that we find essentially the same results for spherical KCl particles as for an ensemble of randomly oriented KCl cubes with rounded corners, consistent with the results for NaCl in Ref.~\cite{Berg2012}. The ionizing radiation is modeled as a monochromatic plane wave which is linearly polarized along the $y$-axis, providing a good model for the synchrotron radiation used in the experiment. We assume that the probability of generating an electron at a specific location $\vec{a}$ inside the particle is proportional to the intensity of the internal electric field $\vec{E}(\vec{a})$ at this location (subsection~\ref{model-2}). The electric fields inside and outside the particle are calculated with the discrete dipole approximation (DDA) using the ADDA package~\cite{Yurkin2007,Yurkin2011,Hoekstra1998,Yurkin2013,Tyynela2009}. We tested the accuracy of this approach with the Lorenz-Mie method which provides an analytical solution for spherical particles and with previously published data ~\cite{Bohren1983,Mie1908,Schafer2012,Lee1990,Schafer2011,Deirmendjian1963}.

\begin{figure}
\begin{center}
\includegraphics[width=8.6cm]{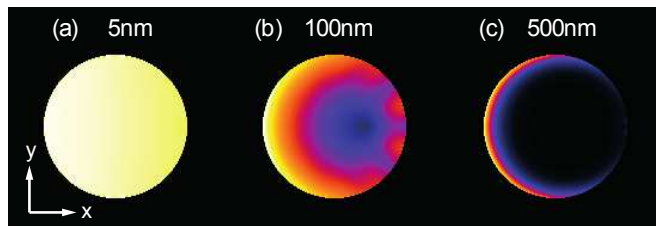}
\end{center}
\caption{(Color online) Calculated intensity patterns of the ionizing radiation inside KCl nanospheres with a radius of 5~nm (panel a), 100~nm (panel b), and 500~nm (panel c), respectively. The calculations are for a photon energy of 10~eV. The light intensity decreases from yellow to red to blue color.}
\label{fig:adda}
\end{figure}

Figure~\ref{fig:adda} shows the squared amplitude of the internal field $|\vec{E}(\vec{a})|^2$, i.e.\ the light intensity, in the $x$-$y$~plane through the center of a spherical particle with a radius of 5~nm, 100~nm, and 500~nm, respectively, at a photon energy of 10~eV. The light intensity inside the smallest particles is homogeneous. Only a very minor decrease in intensity in the propagation direction of the light (positive $x$~direction) is visible, which is due to minor absorptions in the particle. The largest particle absorbs the light almost completely already in the outermost surface layer of the particle that faces the incoming radiation. There is no light that penetrates into the particle's core. For intermediate sizes, here a 100~nm particle, the light intensity inside the particle is strongly structured with regions of low and high intensity. As mentioned, these size-dependent intensity patterns together with the influence of the size-dependent curvature on the electron emission determine the anisotropy in the photoelectron images.

\subsection{Electron emission model\label{model-2}}

We use a simple three-step model for the description of the electron emission~\cite{Abbati1974}: The first step is the generation of the photoelectron at location $\vec{a}$ inside the particle. Once formed, it is randomly scattered inside the aerosol particle (second step) until it reaches the surface and is emitted from the particle into vacuum (third step). For the first step, we assume that the probability $p(\vec{a},\theta)$ of generating an electron at location $\vec{a}$ inside the particle with momentum in direction $\theta$ is proportional to the light intensity (subsection~\ref{model-1}) and the classical dipole emission probability, i.e.\ 
\begin{equation}\label{eq:proba}
p(\vec{a},\theta) \propto |\vec{E}(\vec{a})|^2 \cos^2(\theta)
\end{equation}
where $\vec{E}(\vec{a})$ is the internal electric field at location $\vec{a}$ and $\theta$ is the angle between the directions of the internal electric field vector $\vec{E}(\vec{a})$ and the initial velocity vector $\vec{v}{_\text{initial}}(\vec{a})$ of the electron. The initial electron KEs $E_{K,\text{initial}}$ are chosen randomly from a Gaussian distribution modeled after the photoelectron spectrum of gas phase monomer KCl~\cite{Potts1977} in order not to include scattering \emph{a priori}. For photoelectron spectra recorded at photon energy $h\nu$, they fulfill the condition
\begin{equation}\label{eq:init-ke}
0 < E_{K,\text{initial}} \leq h\nu-{E_g}
\end{equation}
where $E_g$ is the band gap, for which we use a value of 8.4~eV~\cite{Ejiri1994}. The Gaussian is centered at $h\nu-{E_g}-\Delta$, where $\Delta=0.34~\text{eV}$ is the half width at half maximum~\cite{Potts1977}. We have found that the exact shape of the curve and the exact initial KE distribution does not significantly influence the asymmetry parameters $\alpha$, $\beta_{\text{backward}}$, and $\beta_{\text{forward}}$.

The second step, the migration of the electron within the particle, is modeled as a random walk with the inelastic MFP and the elastic MFP as step lengths for inelastic and elastic scattering events, respectively. A random walk corresponds to isotropic scattering of the photoelectrons in the particle. This is probably a good assumption here because the directionality of scattering in bulk KCl is weak~\cite{Akkerman1994} and because our KCl particles are likely polycrystalline. The probabilities of inelastic scattering events $p_{\text{I}}$ and elastic scattering events $p_{\text{E}}$ is calculated as
\begin{equation}\label{eq:proba-inel-el}
p_{\text{I}}=1/(s+1) \textrm{ and } p_{\text{E}}=s/(s+1) \textrm{ .}
\end{equation}
where $s$ is the ratio of the inelastic and the elastic MFP. We use $s=0.1$ in this work consistent with Ref.~\cite{Akkerman1994}. As both elastic and inelastic scattering events are isotropic in our model, the value of $s$ does not significantly influence the anisotropy parameters $\alpha$, $\beta_{\text{backward}}$, and $\beta_{\text{forward}}$. Their values for a 100~nm radius particle, for example, vary by less than 1\% if $s$ is varied between 0.1 and 10. Note that scattering of low energy electrons in solids, in particular elastic scattering, is far from being understood~\cite{Akkerman1994}. For the mean energy loss per inelastic scattering event, the energy loss parameter~$\varepsilon$, we use 
\begin{equation}\label{eq:epsilon}
\varepsilon = 10 \text{ meV at 300~K.}
\end{equation}
This corresponds to the value 9~meV at 270~K from Ref.~\cite{Akkerman1994}, which has been corrected to account for the higher temperature of about 300~K in our experiments. For this correction, the temperature dependence of the mean phonon energy from Ref.~\cite{Gail1980} is used. Again, we can show that the exact value of $\varepsilon$ has only a minor influence on the asymmetry parameters. However, it will significantly affect the electron yield. Since we did not perform yield measurements in the present work, we use 10~meV throughout the paper.

For the third step, the emission of the electron at the particle surface into vacuum, the electron must fulfill the classical momentum threshold condition for escape 
\begin{equation}\label{eq:escape}
\tau \leq \arccos \sqrt{E_A/E_{K,\text{surface}}}
\end{equation}
where $\tau$ is the angle between the electron velocity vector and the surface normal and $E_{K,\text{surface}}$ is the electron KE~\cite{Shimizu1992}. Both values are taken just before ejection; i.e.\ at the surface inside the particle. We use a value of $E_A=0.5~\text{eV}$~\cite{Ejiri1994} and subtract it from the normal component of the electron KE just after ejection into vacuum. For every ejected electron, we store its velocity vector and its KE resulting in a 3D electron emission pattern. The corresponding velocity vector and the corresponding KE in the vacuum just after emission are referred to as $\vec{v}_{\text{vacuum}}$ and $E_K$, respectively.

\subsection{Electron projection, anisotropy parameters, photoelectron yield\label{model-3}}

Experimentally, it is very challenging to record this 3D electron emission pattern. Instead, we use velocity map imaging (VMI)~\cite{Eppink1997} in our experiments (section~\ref{exp}), which projects the emitted 3D electron distribution onto a 2D position sensitive electron detector with the help of an electrostatic lens system. The recorded 2D photoelectron images are referred to as "experimental raw images" in the following. To model the experimental projection, we map each electron onto the detector plane with $b e^{-i\Theta}$, where $b\propto\sqrt{E_K-E_{K,z}}$ and $\Theta$ is the azimuthal angle in the $x$-$y$ plane with respect to the $y$ axis. The corresponding photoelectron images are referred to as "calculated raw images".

\begin{figure}
\begin{center}
\includegraphics[width=8.6cm]{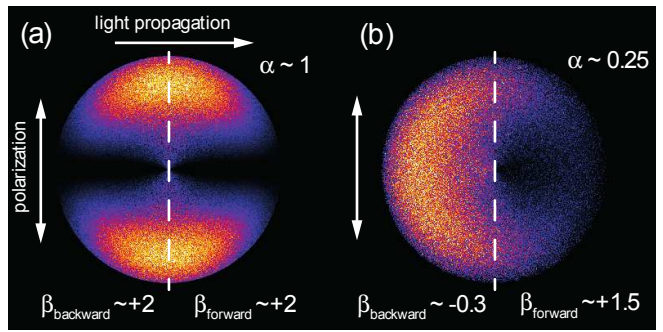}
\end{center}
\caption{(Color online) Example photoelectron images with different anisotropy parameters $\alpha$, $\beta_{\text{backward}}$, and $\beta_{\text{forward}}$. The panels a and b are representative for the situation found for small and large particles, respectively. The backward and forward plane is the left and the right half-plane, respectively. The electron intensity decreases from yellow to red to blue color.}
\label{fig:params}
\end{figure}

To describe the angular dependence of the electron emission patterns, we extract from the raw images three anisotropy parameters, denoted as $\alpha$, $\beta_{\text{backward}}$, and $\beta_{\text{forward}}$. Note that in VMI the 3D emission pattern can be retrieved from a reconstruction of the 2D raw image only if the electron distribution has an axis of cylindrical symmetry parallel to the detector plane ~\cite{Dribinski2002}. In our case, however, the light attenuation inside larger particles (Fig.~\ref{fig:adda}) breaks this symmetry so that reconstruction is not possible. Therefore, we determine the anisotropy parameters directly from the calculated or experimental raw image. For their definition, we split the image plane into two half-planes with $x < 0$ and $x > 0$ referred to as "backward plane" and "forward plane", respectively (Fig.~\ref{fig:params}). The anisotropy parameter $\alpha$ describes the anisotropy of the photoelectron image with respect to the propagation direction of the light. It is defined as in Refs.~\cite{Berg2012,Wilson2007} as
\begin{equation}\label{eq:alpha}
\alpha=I_{x+}/I_{x-}
\end{equation}
where $I_{x+}$ and $I_{x-}$ are the number of photoelectrons projected onto the $x > 0$ and $x < 0$ half-planes, respectively. For gas phase molecules, $\alpha=1$ because there is no anisotropy in the propagation direction of the light for these samples. For KCl aerosol particles, by contrast, one expects $\alpha$ values between 1 and 0 as a result of the attenuation of the light by absorption inside the particle (Fig.~\ref{fig:adda}). For very small aerosol particles, $\alpha$ is still close to 1 because light absorption is minor here (panel~a of Fig.~\ref{fig:params}). With increasing particle size, $\alpha$ decreases continuously. An example for an image with $\alpha = 0.25$ is provided in the panel~b of Fig.~\ref{fig:params}.

In VMI photoemission imaging, the anisotropy is usually described by the anisotropy parameter $\beta$, which is defined through $I(\phi) \propto [1+\beta/2(3 \cos^2 \phi-1)]$ ~\cite{Cooper1968,Zare1988,Melko2013}. The angle $\phi$ is the polar angle with respect to the polarization direction of the light ($y$ direction) and $I(\phi)$ is the number of photons detected in angle $\phi$ in the reconstructed image (center slice image). Here, we keep the same definition for $\beta$, except that we determine it directly from the raw image and separately as $\beta_{\text{backward}}$ in the backward half-plane and as $\beta_{\text{forward}}$ in the forward half-plane (Eq.~\ref{eq:beta}). As a consequence, these $\beta$ parameters do not have the same physical meaning as usual. Rather they are phenomenological parameters used to characterize the anisotropy associated with the polarization direction of the light. We define them by
\begin{equation}\label{eq:beta}
\begin{array}{ll}
I(\phi^-) \propto [1+\beta{_{\text{backward}}/2}(3 \cos^2 \phi^- -1)] \qquad \text{and}\\
I(\phi^+) \propto [1+\beta{_{\text{forward}}/2}(3 \cos^2 \phi^+ -1)]
\end{array}
\end{equation}
where $\phi^-$ and $\phi^+$ are the polar angles in the raw image with respect to the polarization direction of the light ($y$ direction) in the negative ($x < 0$) and positive ($x > 0$) half-planes of the image, respectively, and $I(\phi^-)$ and $I(\phi^-)$ are the numbers of photoelectrons detected in directions $\phi^-$ and $\phi^+$, respectively. Examples of raw images with different $\beta$ parameters are shown in Fig.~\ref{fig:params}. Although the three asymmetry parameters are not completely independent, they all contain complementary information which is useful for the extraction of electron MFPs.

Lastly, we define the total photoemission yield $\gamma$ as a fourth parameter to characterize the electron emission. The parameter $\gamma$ is the ratio of the number of photoelectrons that escape from the particle $q_{\text{out}}$ and the number of absorbed photons $cE_{\text{total}}^2$, i.e.\ 
\begin{equation}\label{eq:gamma}
\gamma=q_{\text{out}}/cE_{\text{total}}^2=\sum_a{p_a}E_i^2{(\vec a)}/E_{\text{total}}^2
\end{equation}
where $q_{\text{out}}=c\sum_a{p_a}E_i^2{(\vec a)}$ and $c$ is a measure of the absorption cross section. The index $a$ runs over all photoelectrons that are formed at location $\vec{a}$ in the particle. $E_i^2{(\vec a)}$ is the local field intensity at location $\vec{a}$ and $E_{\text{total}}^2=\sum_a{E_i^2(\vec a)}$. We assume the probability of generating an electron at location $\vec{a}$ to be proportional to the local field intensity at this position and $p_a\in\{0,1\}$. $p_a$ is~1 if the electron generated at location $\vec{a}$ is finally emitted from the particle and it is~0 for electrons that are generated in the particle but never emitted from the particle.

\section{Velocity map imaging experiment\label{exp}}

We have performed preliminary photoemission experiments on KCl aerosol particles with radii of 37.5, 50, 75, 100, 125, and 150~nm at photon energies of 10, 11.4, and 12.4~eV at the VUV beamline of the Swiss Light Source at the Paul Scherrer Institut. The KCl aerosol particles are generated in an atomizer (TSI Inc., model 3076) and dried after the exit of the atomizer with a home-made diffusion dryer. After size-selection in a scanning mobility particle sizer (SMPS, TSI Inc., model 3080), they enter the photoelectron spectrometer through a home-made aerodynamic lens~\cite{Ziemann1998,Signorell2014,Yoder2013}. Typical particle densities after size selection vary from $0.6\cdot10^5$~to~$3.8\cdot10^5$ particles/cm$^{3}$. VUV synchrotron radiation is used for single-photon ionization of the aerosol particles. The generated photoelectrons are projected onto a 2D electron imaging detector (Roentdek, 40~mm diameter) with a home-made VMI lens system~\cite{Signorell2014,Yoder2013,West2013}. Typical acquisition times are on the order of several hours per image.

High background signal and time constraints at the beam line limited the quality of the images and made measurement on smaller particles impossible, while fluctuations in the photon flux made photoelectron yields unreliable. In the uncertainty estimates given in section~\ref{resu-comparison}, we have tried to account for the most significant sources of error. Hence, the experimental data we present are only preliminary and could be significantly improved and extended by further optimization of the experimental setup by reduction of background noise and increased photon flux stability. However, section~\ref{resu-comparison} demonstrates that even these preliminary experimental data strongly support our proof of principle study.

\section{Results and discussion\label{resu}}

\subsection{Model predictions\label{resu-model}}

\begin{figure}
\begin{center}
\includegraphics[width=8.6cm]{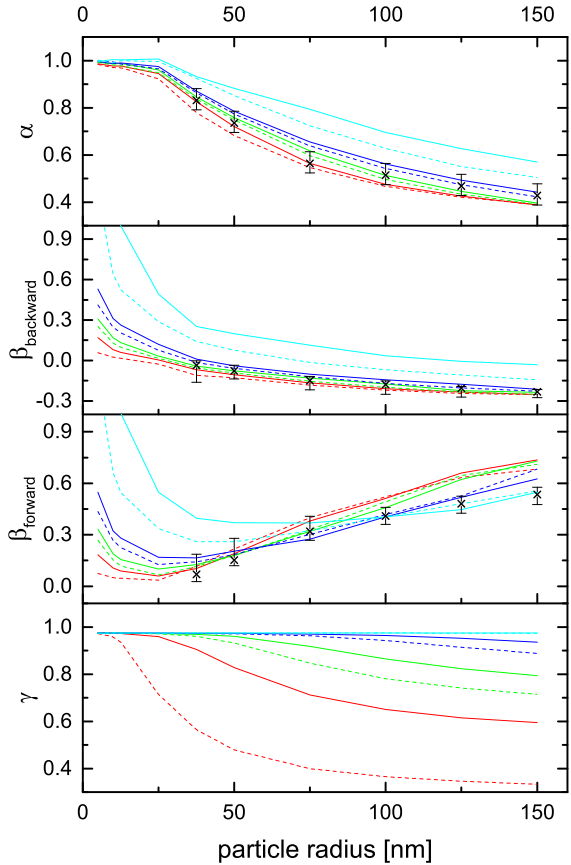}
\end{center}
\caption{(Color online) The lines are simulated anisotropy parameters $\alpha$, $\beta_{\text{backward}}$, and $\beta_{\text{forward}}$ and electron yield $\gamma$ for inelastic MPFs of 0.2 (dashed red), 0.5 (solid red), 0.75 (dashed green), 1 (solid green), 1.5 (dashed blue), 2.5 (solid blue), 5 (dashed cyan) and 10~nm (solid cyan). The experimental data are shown as black crosses with error bars. All data are for KCl particles photoionized at 10~eV photon energy. Please note that the values of $\beta_{\text{forward}}$ for different MFPs cross each other.}
\label{fig:10eV_mfp}
\end{figure}

Figure~\ref{fig:10eV_mfp} shows the evolution of the anisotropy parameters and the total electron yield as a function of the size of the KCl particles as predicted by the model. The different lines represent values for eight different inelastic MFPs from 0.2 to 10~nm (see caption of Fig.~\ref{fig:10eV_mfp}). For these simulations, we assume at first that the inelastic MFP is constant during the random walk, i.e.\ that it does not depend on the KE of the electron. Figure~\ref{fig:10eV_mfp} reveals that all four parameters depend on the inelastic MFP, but in different ways. The parameter $\alpha$ depends only weakly on the MFP for particle radii below about 25~nm because the light intensity in these particles is more or less homogeneous (Fig.~\ref{fig:adda}) and the number of scattering events is low to moderate. For larger sizes, however, the values of $\alpha$ become more clearly distinguishable for different inelastic MFPs, which is mainly a consequence of the inhomogeneous light intensity inside the aerosol particles (Fig.~\ref{fig:adda}). Moreover, $\gamma$ shows a similar trend. It is less sensitive to the MFP for smaller particles than for larger ones, for which it shows a distinct dependence on the MFP. Just the opposite behavior is observed for the two $\beta$ parameters. In contrast to $\alpha$ and $\gamma$, $\beta_{\text{backward}}$ and $\beta_{\text{forward}}$ are clearly more sensitive to the MFP for particle radii below about 50~nm than for larger particles. The complementary and characteristic dependence of the four parameters on the inelastic MFP is useful in that if one covers a broad aerosol particle size range one obtains comprehensive information on the MFP. This, in turn, may allow one to determine accurate values for MFP using aerosol measurements.

\begin{figure}
\begin{center}
\includegraphics[width=8.6cm]{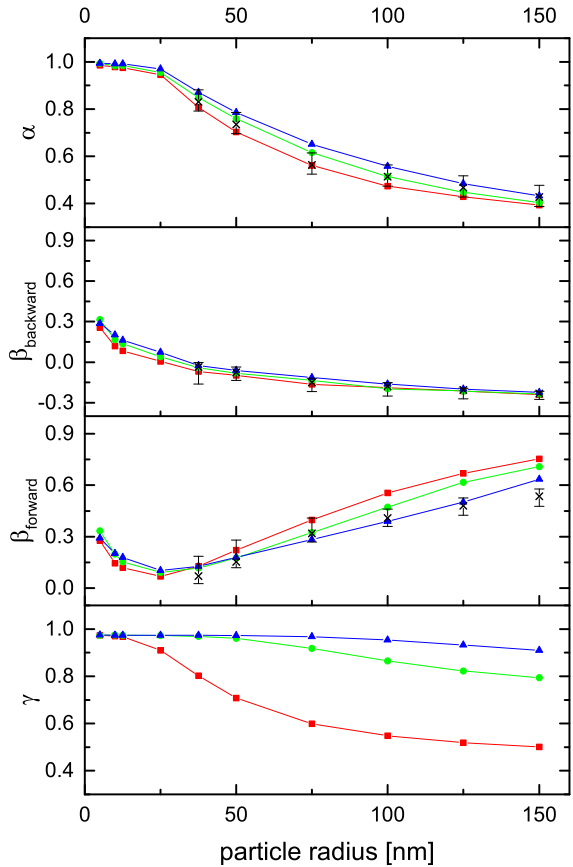}
\end{center}
\caption{(Color online) Anisotropy parameters $\alpha$, $\beta_{\text{forward}}$, and $\beta_{\text{backward}}$ and electron yield $\gamma$ simulated for a constant inelastic MFP of 1~nm (green dots), an energy dependent inelastic MFP representing the interaction with the acoustic mode (blue triangles), and an energy dependent inelastic MFP representing the interaction with the optical mode (red squares)~\cite{Akkerman1994}. The experimental data are shown as black crosses with error bars. All data are for KCl particles photoionized at 10~eV photon energy.}
\label{fig:en-dep}
\end{figure}

For the simulations in Fig.~\ref{fig:10eV_mfp}, we have assumed that the inelastic MFP remains constant throughout the random walk of the electron in the particle. In reality, however, the electron MFP can pronouncedly depend on the KE of the electron, which then changes during the random walk. From the observed dependence of the four parameters on the MFP in Fig.~\ref{fig:10eV_mfp}, one would expect to find noticeable effects on these parameters when the energy dependence of the inelastic MFP is included. Fig.~\ref{fig:en-dep} confirms this expectation. This figure shows the dependence of the four parameters on the inelastic MFP as a function of particle size for three different cases: a constant inelastic MFP of 1~nm (identical to the solid green line in Fig.~\ref{fig:10eV_mfp}) and two inelastic MFPs that depend in different ways on the KE of the electron. The energy dependences of the latter are derived from the calculated MFPs for acoustic phonon-electron interaction and optical phonon-electron interaction provided in Fig.~1~(c) of Ref.~\cite{Akkerman1994}. For both types of interaction, the energy dependence is to a good approximation linear in the region that is of interest here. We have approximated them by two straight lines with a slope of +0.8~nm/eV and -0.8~nm/eV, respectively, both of them sporting the value of 1~nm MFP at 1~eV electron KE. The influence of the energy dependence of the MFPs on the anisotropy parameters in Fig.~\ref{fig:en-dep} is in general moderate in particular for the smaller particles. An exception is $\beta_{\text{forward}}$ for larger particles, which exhibits a pronounced energy dependence in this size range. The parameter $\beta_{\text{forward}}$ is mainly determined from electrons that originate from the particle half-sphere in which for larger particles the intensity pattern of the light varies sensitively with particle size (Figs.~\ref{fig:adda} and \ref{fig:params}). As a consequence, $\beta_{\text{forward}}$ is here also very sensitive to the inelastic MFP. The strong dependence of the electron yield $\gamma$ on the MFP is not a surprise since it scales with the MFP (Eq.~\ref{eq:gamma}).

The dependence of the four parameters on the refractive index $n+ ik$ is illustrated in Fig.~\ref{fig:50nm_9-11eV}, for the example of a 50~nm KCl particle and for a constant inelastic MFP of 1~nm, again neglecting the dependence of the MFP from the electron KE as in Fig.~\ref{fig:10eV_mfp}. The calculations are for the KCl refractive index data between 9 and 13.5~eV displayed in the two top traces in Fig.~\ref{fig:50nm_9-11eV}. The green dots represent calculations for which only the refractive index and the wavelengths of the light vary while all other parameters are kept constant. In particular, the initial KE distribution $E_{K,\text{initial}}$ is not varied according to the incident photon energy (section~\ref{model-2}). We chose a constant $E_{K,\text{initial}}$ distribution as calculated for a photon energy of 10~eV (section~\ref{model-2}). The variations in $\alpha$, $\beta_{\text{backward}}$, $\beta_{\text{forward}}$ in Fig.~\ref{fig:50nm_9-11eV} follow more or less the variations of the refractive index. The changes in the absolute values of $\alpha$, $\beta_{\text{backward}}$, and $\beta_{\text{forward}}$ are moderate to pronounced. The absolute value of $\gamma$, by contrast, is hardly influenced by the refractive index. A similar behavior is also observed for other particle sizes (data not shown). For the black squares in Fig.~\ref{fig:50nm_9-11eV}, the $E_{K,\text{initial}}$ distribution is varied in addition to the refractive index and wavelength. The $E_{K,\text{initial}}$ distribution at each photon energy is calculated as described in section~\ref{model-2}. The comparison with the green dots (dashed lines) reveals that the additional influence of the initial kinetic energy is not very pronounced, except for $\gamma$ at very low photon energies. Fig.~\ref{fig:50nm_9-11eV} reveals that the quality of refractive index data is important for the determination of accurate MFPs from anisotropy parameters.

\subsection{Comparison with experiments\label{resu-comparison}}

\begin{figure}
\begin{center}
\includegraphics[width=8.6cm]{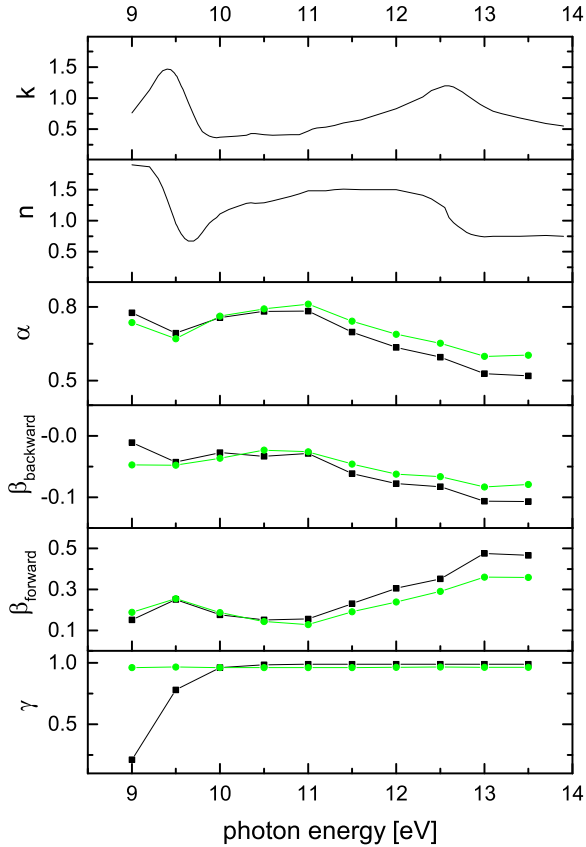}
\end{center}
\caption{(Color online) The two top traces show the refractive index data $n+ ik$ from Ref.~{\cite{Roessler1968}} as a function of the photon energy. The four bottom traces show the anisotropy parameters $\alpha$, $\beta_{\text{backward}}$, and $\beta_{\text{forward}}$ and the total electron yield $\gamma$ as a function of the photon energy. All calculations are for a constant inelastic MFP of 1~nm and a KCl particle radius of 50~nm. For the green dots, it is assumed that only the refractive index varies with photon energy. For the black squares it is assumed that the refractive index and the initial kinetic energy distribution vary with photon energy. Note that the lines in the four lower panels are merely meant to guide the eye.}
\label{fig:50nm_9-11eV}
\end{figure}

The modeling results clearly reveal the principal advantage of aerosol studies over other methods for the determination of accurate electron MFPs . The advantage lies in the characteristic dependence of the anisotropy parameters on the EMFP as a function of aerosol particle size (Fig.~\ref{fig:10eV_mfp}). If a broad size range is covered the complementary information contained in the three anisotropy parameters allows one to extract detailed information on the EMFP. Given that high quality experimental data are available, one should be able to extract accurate EMFPs from a fit of calculated to experimental data, including information on the energy dependence of the EMFP (Figs.~\ref{fig:en-dep} and~\ref{fig:50nm_9-11eV}).

To complete our proof of principle study, this section provides a comparison of the modeling results with preliminary experimental data for KCl aerosol particles. As mentioned in section~\ref{exp}, the quality of the experimental data is limited and it was thus not possible to extract reliable data for the electron yield $\gamma$ or to record meaningful data for very small aerosol particles. We have recorded photoemission images at photon energies of 10, 11.4, and 12.4~eV for six different particle sizes. As examples, we present mainly the data for 10~eV photon energy and only a summary for 11.4~eV photon energy. Because of the low signal to noise of the measurements recorded at 12.4~eV and overlapping signals from background gas phase water we do not provide any data for this photon energy. In view of the limited quality of the experimental data, we do not attempt to extract any energy dependent information on the inelastic MFP (Fig.~\ref{fig:en-dep}), but compare it first only with the model predictions for constant MFPs. This comparison is shown in Fig.~\ref{fig:10eV_mfp}. Despite the preliminary character of the experimental data, both the experimental and the calculated anisotropy parameters show the same trends with increasing aerosol particle size. The values of $\alpha$ and $\beta_{\text{backward}}$ decrease continuously with increasing particle size while $\beta_{\text{forward}}$ increases with increasing size. All values of $\alpha$ and $\beta_{\text{backward}}$ are consistent with a short MFP. The comparison of these data with the model predictions results in an average inelastic MFP of about 1~nm at an average electron energy of about 1~eV. A similar analysis of $\alpha$ and $\beta_{\text{backward}}$ for the measurement at a photon energy of 11.4~eV results in an average inelastic MFP of about 0.5~nm at an average electron energy of about 2.5~eV. Both values are in accordance with the theoretical prediction of Ref.~\cite{Akkerman1994}. Note that our inelastic MFP corresponds to the dominating inelastic MFP, i.e.\ the shorter one in Ref.~\cite{Akkerman1994}. On the basis of the present data, we are unable to judge how significant the agreement with Ref.~\cite{Akkerman1994} is. It should be noted in this context that Akkerman et al.\ use a similar model as the present work; i.e.\ they also treat the electrons as classical particles that move along stochastic trajectories. The main difference to our model lies in the MFPs. They use a quantum mechanical model to calculate the MFPs while we extract them from a fit to experimental data. At least, the comparison suggests that the modeling approach chosen allows for a meaningful analysis of the experimental data.

\begin{figure}
\begin{center}
\includegraphics[width=8.6cm]{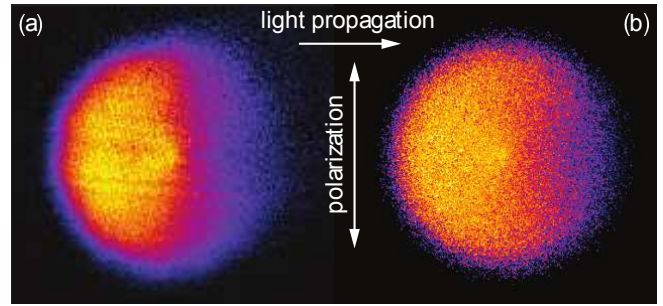}
\end{center}
\caption{(Color online) Experimental (panel a) and calculated (panel b) raw photoelectron image of a KCl particle of 150~nm radius. The particles are ionized with light of 10~eV photon energy.}
\label{fig:exp-vs-sim}
\end{figure} 

In the above estimates of the MFPs, we have not used $\beta_{\text{forward}}$. Fig.~\ref{fig:10eV_mfp} shows that $\beta_{\text{forward}}$ is only consistent with an inelastic MFP of about 1~nm for the small particle size, but not above about 100~nm. The same phenomenon is found for the data for 11.4~eV photon energy (not shown). At this point, we can offer two explanations. The inconsistency of $\beta_{\text{forward}}$ observed for larger particle sizes could be a consequence of the low signal to noise in the photoelectron images. Figure~\ref{fig:exp-vs-sim} shows experimental and calculated photoelectron images for a 150~nm KCl particle. The parameter $\beta_{\text{forward}}$ is defined in the half-plane in which the number of detected electrons is low for larger particles (Fig.~\ref {fig:params}) because they originate from the particle hemisphere that is only weakly illuminated (Fig.~\ref {fig:adda}). Therefore, the signal to noise ratio is lower and background contributions in the experimental image are higher in the half-plane which is used to determine $\beta_{\text{forward}}$ compared with the one that is used for $\beta_{\text{backward}}$. This can result in ill-defined values for $\beta_{\text{forward}}$. The inconsistency of $\beta_{\text{forward}}$ for larger particles might also originate from the energy dependence of the inelastic MFP. As discussed, Fig.~\ref{fig:en-dep} shows that $\beta_{\text{forward}}$ is more sensitive to this energy dependence compared with $\beta_{\text{backward}}$ and $\alpha$. The comparison of experimental and calculated data in this figure also reveals that the energy dependence of the inelastic MFP can easily account for the apparent inconsistency of $\beta_{\text{forward}}$. The data for the interaction with the acoustic mode and a 1~nm inelastic MFP do no longer show any systematic deviation of $\beta_{\text{forward}}$ for larger particle sizes. For an energy dependence of the inelastic MFP according to the acoustic mode, all three anisotropy parameters provide consistent results for the inelastic MFP. More accurate measurements will resolve which of these effects dominates, background contributions or the energy dependence of the inelastic MFP. But already the preliminary experimental data discussed here provide a convincing proof of principle for the new aerosol approach to determine MFPs of slow electrons.

\section{Conclusions\label{concl}}
This proof of principle study shows that angle-resolved aerosol photoelectron imaging opens up a new and potentially more accurate way to determine electron MFPs for low KE electrons in solids and liquids, provided that accurate experimental data and a realistic model are available. The advantage of the aerosol approach arises from the characteristic information on the MFP that is contained in the anisotropy of the photoelectrons. The variation of the particle size - and thus the variation in curvature - results in characteristic modifications of the light absorption and the electron transport inside the aerosol particles and thus of the angular dependence of the photoelectrons. Our modeling results clearly demonstrate that the use of the anisotropy information in principle allows one to extract detailed information on the MFPs from fits of simulated photoelectron images to experimental ones.

Important for the determination of accurate MFPs is the quality of the experimental photoelectron images. Crucial are a good signal to noise ratio, low background contributions, a quality electron imaging system, and stable particle and photon fluxes. The first two factors strongly influence the quality of the anisotropy data. The third factor is crucial to obtain reliable data for the total electron yield. For the present study, we use a velocity map imaging photoelectron spectrometer. We do not fit simulated photoemission images directly to experimental ones. Instead, we first define four parameters (the electron yield and three anisotropy parameters) from the images, which we then use for the comparison between simulation and experiment. A major reason for this choice is the quality of the preliminary experimental data. In future studies, it is planned to directly fit whole images in order to exploit their full information content. The electron MFP is a model-dependent quantity. Here, we used a simple conventional model as a start. The incorporation of more sophisticated models might further increase the utility of the aerosol approach ~\cite{Burtscher1982,Akkerman1996,Hansson1998,Akkerman1994,DiStefano1973}.

The aerosol approach is particularly attractive for very low-kinetic energy electrons and for liquids for which experimental data on MFPs are urgently needed. MFPs of low-kinetic energy electrons are important in many different fields ranging from astrophysics to radiotherapy, and liquid phase chemistry. We have performed first experiments for various liquids and are currently applying the aerosol approach to these systems.

\begin{acknowledgments}
We would like to thank Egor Chasovskikh for the design of the aerodynamic lens and David Stapfer from our LPC mechanical shop for the fabrication of the aerosol dryer, the aerodynamic lens, and the VMI optics. The photoemission measurements were performed on the VUV beamline at the Swiss Light Source, Paul Scherrer Institut, Villigen, Switzerland. We are grateful to Andr\'as B\H{o}di at the Swiss Light Source for his support during the experiments. Financial support from the Swiss National Science Foundation (SNSF grant No.~$200021\_146368/1$) and the ETH Z\"urich is acknowledged. 
\end{acknowledgments}



%

\end{document}